\documentclass[11pt]{article}
\pdfoutput=1

\usepackage{fullpage}
\usepackage{amsmath}
\usepackage{amssymb}
\usepackage{setspace}
\usepackage{bbm}
\usepackage{dsfont}
\usepackage{graphics}

\usepackage{color}

\usepackage[colorlinks=true]{hyperref} 
\hypersetup{
    bookmarks=true,         
    unicode=false,          
    pdftoolbar=true,        
    pdfmenubar=true,        
    pdffitwindow=false,     
    pdfstartview={FitH},    
    pdftitle={My title},    
    pdfauthor={Author},     
    pdfsubject={Subject},   
    pdfcreator={Creator},   
    pdfproducer={Producer}, 
    pdfkeywords={keyword1} {key2} {key3}, 
    pdfnewwindow=true,      
    colorlinks=true,       
    linkcolor=blue,          
    citecolor=red,        
    filecolor=magenta,      
    urlcolor=cyan           
}

\usepackage[normalem]{ulem}
\usepackage{amsmath}
\usepackage{enumerate}
\usepackage{amsfonts}
\usepackage{yfonts}

\usepackage{subfigure}
\usepackage{psfrag}

\usepackage{epsfig}
\usepackage[latin1]{inputenc}
\usepackage{float}
\usepackage{graphicx}
\usepackage{cancel}
\usepackage{mathrsfs}
\usepackage{amssymb}
\usepackage{amsfonts}
\usepackage{amsmath}
\usepackage{slashed}
\usepackage[font=small,labelfont=bf]{caption}

\newcommand \tr {\mbox{{\bf Tr}}}

\usepackage{graphicx}
\usepackage{bm}

\newcommand{\be}{\begin{equation}}
\newcommand{\ee}{\end{equation}}
\newcommand{\bes}{\begin{equation*}}
\newcommand{\ees}{\end{equation*}}
\newcommand{\bea}{\begin{eqnarray}}
\newcommand{\eea}{\end{eqnarray}}
\newcommand{\beas}{\begin{eqnarray*}}
\newcommand{\eeas}{\end{eqnarray*}}

\newcommand{\bmat}{\begin{bmatrix}}
\newcommand{\emat}{\end{bmatrix}}

\def\tr{{\rm tr}}

\begin{document}

\numberwithin{equation}{section}
{
\begin{titlepage}
\begin{center}

\hfill \\
\hfill \\
\vskip 0.75in

{First Law of Entanglement Entropy vs Black Hole Entropy in an Anomalous Theory}\\

\vskip 0.3in

{ Long Cheng${}^{a,b}$, Ling-Yan Hung${}^{a,b,c}$, Si-Nong Liu${}^{a,b}$, Hong-Zhe Zhou${}^{a,b}$ }\\
\vskip 0.3in
{\it ${}^a$ State Key Laboratory of Surface Physics and Department of Physics, Fudan University,
\\ 220 Handan Road, 200433 Shanghai, China}\vskip .5mm
{\it ${}^{b}$ Center for Field Theory and Particle Physics, Fudan University, \\
220 Handan Road, 200433 Shanghai, China} \vskip .5mm
{\it ${}^{c}$ Collaborative Innovation Center of Advanced  Microstructures,
Fudan University,\\ 220 Handan Road,  200433 Shanghai, China.}

\end{center}

\vskip 0.35in

\begin{center} {\bf Abstract }\\
  \end{center}
In this note we explore the validity of the first law of entanglement entropy in the context of the topologically massive gravity (TMG). We found that the variation of the holographic entanglement entropy under perturbation from the pure AdS background satisfies the first law upon imposing the bulk equations of motion in a given time slice, despite the appearance of instabilities in the bulk for generic gravitationl Chern-Simons coupling $\mu$. The black hole entropy is different from the holographic entanglement entropy in a general boosted frame. This discrepancy however do not affect the entanglement first law.

\vfill

\noindent \today

\end{titlepage}
}

\newpage

\tableofcontents
\newpage

\section{Introduction }
There is now overwhelming evidence that the properties of quantum entanglement of a quantum field theory is profoundly ingrained in the structure of gravity, which should be expected if the gauge gravity correspondence should be true. One very beautiful example illustrating such a connection is the study of relative entropy in a field theory, whose gravity dual is found to be directly related to the Einstein equation \cite{Blanco:2013joa, Lashkari:2013koa, Faulkner:2013ica}.
In fact, imposing that the relative entropy is positive definite, which follows from unitarity in the dual field theory, one can derive the linearized version of Einstein equation, and which is subsequently generalized to generic higher derivative theories \cite{Lashkari:2013koa, Faulkner:2013ica}. It is also argued that this should imply also the non-linear version of Einstein equation \cite{Swingle:2014uza}. 

These insights are positive affirmation that when unitarity is assured in the theory, the gravity theory correctly recovers a positive relative entropy. An interesting test therefore is the opposite: if unitarity is lost, in which the gravity and also its gauge theory dual reports instablilites, is the relative entropy sensitive to 
these pathologies? Moreover, there is recently a fresh surge of activity towards understanding systems with anomaly, partly stimulated by their relevance to the understanding of symmetry protected topological phases of matter \cite{CGLW,Wen:2013oza,Kong:2014qka}. The manifestation of gravitational anomaly in entanglement entropy \cite{Iqbal:2015vka}, and their holographic duals which involve Chern-Simons terms, are recently discussed for example in \cite{Castro:2014tta,Azeyanagi:2015uoa, Guo:2015uqa}. 

We would therefore like to explore these issues in a well-known example in 2+1 d gravity, known to be dual to a CFT suffering from gravitational anomaly, and at the same time exhibits instabilities even in the pure AdS background -- namely the topological massive gravity (TMG). 
The action of the theory is given by
\be
S =\frac{S_{EH}}{ {16\pi G_N}}+  \frac{S_{TMG} }{32\pi G_N \mu}= \frac{1}{16\pi G_N}\int d^3x \,\, \sqrt{g}  (R+ \Lambda)  + \frac{1}{32\pi G_N \mu}\int \tr(\Gamma \wedge d \Gamma  + \frac{2}{3} \Gamma \wedge \Gamma \wedge \Gamma).
\ee 
In the following we are loosely using the word TMG to refer to gravitational Chern-Simons term. The entanglement entropy of the theory has been worked out \cite{Castro:2014tta}. It is given by
\be
S_{EE}=\frac{1}{4G_N}\int_C ds \sqrt{g_{\mu\nu}\dot X^\mu \dot X^\nu} +  \frac{1}{\mu}\tilde n. (t^s\nabla_s n) + \frac{1}{\mu}S_{constraint},\label{TMGEE}
\ee where a dot in $\dot X$ denotes derivative along the (unit) tangent vector $t^\mu$ of $C$, and $\tilde n$ and $n$ are the two normals of the trajectory $C$.   The constraints are implemented by introducing $S_{constraint}$ which is given by \be  \label{const}
S_{constraint} = \int_C ds [\lambda_1 n.\tilde n + \lambda_2 n.t + \lambda_3 \tilde n.t + \lambda_4 (n^2 +1) + \lambda_5 (\tilde n^2 -1)].
\ee The analysis of \cite{Li:2008dq} demonstrates that there are negative energy modes away from the chiral point, while precisely at the chiral point, there are still pathological modes \cite{Skenderis:2009nt, Carlip:2008eq} excluded only because they are not falling off quickly enough in the AdS boundary. Including them however correspond to  a non-unitary Log-CFT -- the extra modes correspond to including an operator with negative two point correlation function.  With these multiple knobs that lead to instability, we can test how turning on each of these pathologies could change the holographic relative entropy, and thus testing the sensitivity of the entanglement entropy towards them. 

As we will see, the entanglement first law holds despite the loss of unitarity, and the discrepancy between the black hole entropy and the holographic entanglement entropy. 

The paper is organized as follows:
we begin in section \ref{noether} with the derivation of the black hole entropy as a Noether charge in  TMG theory following \cite{Wald:1993nt} and \cite{Tachikawa:2006sz}.  In section \ref{violate} we compute explicitly the perturbation of the black hole entropy away from the pure AdS background, which can be related to an asymptotic energy via the use of the equations of motion. This is the usual first law of black hole physics. We then demonstrate that connection between the first law of black hole physics and the positivity of relative entropy demonstrated in \cite{Blanco:2013joa, Lashkari:2013koa, Faulkner:2013ica} is \emph{preserved} even in the presence of unstable modes in the bulk.
We conclude in section \ref{conclude}.
 Some details of the calculation is relegated to the appendix.

\section{Black hole entropy functional as a Noether charge in TMG} \label{noether}

A proposal for computing the entanglement entropy of the TMG theory was proposed in \cite{Castro:2014tta}.  It is known that the Wald like entropy derived as a conserved Noether charge exactly coincides with the entanglement entropy when we consider the entanglement entropy of a single interval of the CFT ground state corresponding to the pure AdS background \cite{ Castro:2014tta, Casini:2011kv}. The world-line that gives the holographic entanglement entropy can be mapped to a horizon via a coordinate transformation \cite{Casini:2011kv}. 

Here, we would like to derive the explicit form of the Noether current. The Noether current $J$ is obtained by 
\be
\int d J = \int d\Theta - \delta_\xi \mathcal{L}  + \delta_\xi g_{\mu\nu} \,(\textrm{EOM})^{\mu\nu}.
\ee
The term $d\Theta$ is the total derivative term following from varying the action wrt to the metric and integrating by parts to isolate the equations of motion. In the following, we will focus on the contribution of the TMG term.  We have
\be
\Theta_{TMG}  =  - {\text{tr}}({{\Gamma }} \wedge \delta {{\Gamma }}) + \frac{1}
{2}\delta {g_{\mu \lambda }}{R^{\nu \lambda }}_{\rho \sigma }{{\hat \epsilon }^{\mu \rho \sigma }}{{\hat \epsilon }_{\nu \alpha \beta }}{\text{d}}{x^\alpha } \wedge {\text{d}}{x^\beta },
\label{ThetaTMG}
\end{equation}
which is the total derivative term obtained by varying the metric and subtracting by the equations of motion. For completeness, the TMG contribution to the equations of motion mentioned above is given by
\be
(\textrm{EOM}(TMG))^{\mu\nu}=  \frac{1}
{2} ({\nabla _\lambda }{R^{\nu \lambda }}_{\rho \sigma }{{\hat \epsilon }^{\mu \rho \sigma }} + {\nabla _\lambda }{R^{\mu \lambda }}_{\rho \sigma }{{\hat \epsilon }^{\nu \rho \sigma }}).
\ee
The second term is the variation of the action under diffeomorphism 
\be
\delta x^\mu= \xi^\mu(x), \qquad \delta g_{\mu\nu} = 2\nabla_{(\mu} \xi_{\nu)}.\label{diffeo}
\ee
Under such a transformation, the variation of the Christoffel symbol can be written as
\be
\delta_\xi \Gamma =  dv + [\Gamma, v] + i_\xi R,
\ee
where 
\be
v^\mu_\nu = \nabla_\nu \xi^\mu, \qquad R = d\Gamma + \Gamma \wedge \Gamma,\qquad i_\xi V= \frac{1}{(n-1)!} \xi^\mu V_{\mu \nu_2\cdots \nu_n} dx^{\nu_2}\wedge \cdots dx^{\nu_n}
\ee
and we are treating the Christoffel symbol as a matrix 1-form taking values in the adjoint representation 
of the general linear group $GL(3)$. This makes the discussion of the gravitational Chern -Simons term closely parallel that of the usual non-Abelian Chern-Simons term under a gauge transformation.

\be
\delta_\xi {S}_{TMG} =\int {\text{tr}}\left[ {{\text{d}}({{v}}{\text{d}}{{\Gamma }}) - {\text{d}}({{\Gamma }} \wedge {i_\xi }{{R}})} \right],\ee
where we have made use of the following identity that is valid only in 3-dimensions:
\be
{i_\xi }{{R}} \wedge {{R}} = 0.
\ee

Since the current is conserved up to equations of motion, $J$ is closed up to equations of motion, and one should be able to express
\be
J(\xi) = dQ(\xi) + \textrm{EOM}.
\ee
where the form of the conserved current depends on the choice of diffeomorphism. 
After some tedious manipulations, we find that the current $J$ can be messaged into the following form 
\begin{eqnarray}
J_{TMG}(\xi) &=& {\text{d}}[{\text{tr}}(2{{v\Gamma }} - {i_\xi }{{\Gamma }} \wedge {{\Gamma }})] + \nonumber\\
&& \frac{1}
{2}{\partial _\lambda }[{\xi _\mu }({R^{\nu \mu }}_{\rho \sigma }{{\hat \epsilon }^{\lambda \rho \sigma }} + {R^{\nu \lambda }}_{\rho \sigma }{{\hat \epsilon }^{\mu \rho \sigma }} - {R^{\lambda \mu }}_{\rho \sigma }{{\hat \epsilon }^{\nu \rho \sigma }})]{{\hat \epsilon }_{\nu \alpha \beta }}{\text{d}}{x^\alpha } \wedge {\text{d}}{x^\beta } \hfill 
   + \nonumber\\  
   &&2{\xi _\mu }{({\text{EOM}(TMG)}{\text{}})^{\mu \nu }}\frac{1}
{2}{{\hat \epsilon }_{\nu \alpha \beta }}{\text{d}}{x^\alpha } \wedge {\text{d}}{x^\beta, }  
\end{eqnarray}
and thus,
\be
Q_{TMG}(\xi)= {\text{tr}}(2{{v\Gamma }} - {i_\xi }{{\Gamma }} \wedge {{\Gamma }}) + \frac{1}
{2}{\xi _\mu }({R^{\nu \mu }}_{\rho \sigma }{{\hat \epsilon }^{\lambda \rho \sigma }} + {R^{\nu \lambda }}_{\rho \sigma }{{\hat \epsilon }^{\mu \rho \sigma }} - {R^{\lambda \mu }}_{\rho \sigma }{{\hat \epsilon }^{\nu \rho \sigma }}){{\hat \epsilon }_{\nu \lambda \alpha }}{\text{d}}{x^\alpha }
\ee
We note the factor of 2 appearing in front of the EOM term, which ensures that combining with the Einstein contribution we still recover a current $J$ that is exact up to the overall equations of motion. 

As discussed in \cite{Tachikawa:2006sz}, it is necessary to modify the definition of the conserved charge so that the result is independent of shifts of $J$ by exact terms $J \to J + dO$. Such a definition of the charge is given by
\be
{{Q'}_\xi } \equiv {Q_\xi } - {C_\xi },\ee
where 
\be
\delta {C_\xi } \equiv {i_\xi }\Theta  + {\Sigma _\xi },\qquad \text{d}{\Sigma _\xi } = {\Pi _\xi } - \delta {\Xi _\xi } = 0,\qquad {\text{d}}{\Xi _\xi } \equiv  ({\delta _\xi } - {\mathcal{L}_\xi })L, \qquad {\Pi _\xi } \equiv ({\delta _\xi } - {\mathcal{L}_\xi })\Theta.
\ee
The explicit expressions are presented in the appendix.

We are interested in the variation of the entropy by some generic metric perturbations, corresponding to the change in the entanglement entropy of a single interval when the state is slightly perturbed from the ground state . 
The variation of $Q'$ is given by
\be
\delta Q'_{TMG}(\xi)= {\text{tr}}(2{{U}}\delta {{\Gamma }} + 2{i_\xi }{{\Gamma }} \wedge \delta {{\Gamma }}) + \frac{1}
{2}\delta \{ {\xi _\mu }( - 4{R^\mu }_\beta  + {R^{\nu \lambda }}_{\rho \sigma }{{\hat \epsilon }^{\mu \rho \sigma }}{{\hat \epsilon }_{\nu \lambda \beta }})\} {\text{d}}{x^\beta } 
   - {\xi ^\alpha }\delta {g_{\mu \lambda }}{R^{\nu \lambda }}_{\rho \sigma }{{\hat \epsilon }^{\mu \rho \sigma }}{{\hat \epsilon }_{\nu \alpha \beta }}{\text{d}}{x^\beta },
\ee
where
$U^\beta_\alpha=\partial_\alpha \xi^\beta$.

For completeness, the corresponding result for the Einstein-Hilbert term $S_{EH}$ is given by

\begin{equation}
Q'_{EH}(\xi)=  -{\nabla ^a}{\xi ^b}{\epsilon _{abi}}{\text{d}}{x^i}
\label{3.1}
\end{equation}

Now we take $\xi=\xi_B$, which  is  a killing vector of pure AdS space \cite{Faulkner:2013ica}
\be
\xi_B = -\frac{2\pi}{R} (t-t_0)(z \partial_z + (x-x_0)\partial_x) + \frac{\pi}{R}(R^2 - z^2 - (t-t_0)^2 - (x-x_0)^2))\partial_t
\ee
which vanishes along the entangling surface, since the entangling surface of a single interval in pure AdS space coincides with the AdS Rindler horizon \cite{Casini:2011kv}. 
To set notations we note that in Poincare coordinates the pure AdS metric is given by
\be
ds^2 = \frac{R^2}{z^2} (dz^2 + dx^2 - dt^2).
\ee

The first law in black hole entropy is obtained by making use of the fact that $\delta S = \int_\Sigma \delta Q'$, where $\Sigma$ denotes the entangling surface, or the AdS Rindler horizon, 
and an asymptotic energy $\delta E$ can be defined by evaluating $\delta Q'$ at asymptotic infinity , which in our case, correspond to the AdS boundary $z \to 0$ and along the boundary interval $-R<x<R$. Since the Rindler horizon and the boundary surface on which the asymptotic energy is defined forms a closed surface. We immediately have
\be
\delta S - \delta E =2\pi \int_{M} d\delta Q'=2\pi \int_{M} \textrm{EOM}
\ee
 where $M $ is the region surrounded by the boundary surface and the AdS Rindler horizon, or the entangling surface. The rhs vanishes on-shell, leading to the first law of black hole entropy.
A first law of black hole thermodynamics only turns into a first law of entanglement entropy {\it{if the asymptotic energy $\delta E$ coincides with the entanglement Hamiltonian.}} For a single interval on the slice $t=t_0$, the entanglement Hamiltonian of a CFT in flat space is given by
\be \label{entangleH}
H = 2\pi\int_{-R}^R dx \frac{R^2- x^2}{2R}T_{tt} = \int_{-R}^{R} dx \,\,\xi^\mu_B(t=t_0, z=0)T_{\mu t}.
\ee
We will check in the next section if a first law of black hole entropy should imply a first law of entanglement entropy which is supposedly guaranteed by the positivity of the relative entropy. This is by no means obvious in the presence of unstable modes.

\section{Entanglement entropy first law  }\label{violate}
We will check the first law of entanglement entropy of a single interval in two steps: first in a single time slice, and then  on a  boosted slice. As we shall see, the first law is satisfied, even if the modes demonstrated to be unstable in \cite{Li:2008dq} are turned on.  

\subsection{On a given time slice}
We will check in this section the implication of a black hole first law in a TMG theory.
We will consider the time slice $t=t_0$. The variation of the conserved energy is defined as
\be
\delta E_{\infty} =2\pi \int_{z\to 0, |x|<R, t=t_0}  \delta Q'(\xi_B), \label{dE}
\ee
The above can be evaluated by expanding the perturbed metric near the $z=0$ boundary
in the Fefferman-Graham gauge. 

\subsubsection{For $\mu=1$}

For the specific case $\mu=1$, we have
\be
ds^2= \frac{1}{z^2}(dz^2 + g_{\mu\nu}(x,t,z)dx^\mu dx^\nu),
\ee
where
\be \label{pert}
g_{\mu\nu} = \eta_{\mu\nu} + h_{\mu\nu}(z,x) ,
\ee
and where
\begin{equation}
{\mathbf{h}} = \rho \log \rho \left( {\begin{array}{*{20}{c}}
   b & { - b}  \\
   { - b} & b  \\

 \end{array} } \right) + \rho \left( {\begin{array}{*{20}{c}}
   h & k  \\
   k & h  \\

 \end{array} } \right) + \cdots, \quad \rho = z^2
 \label{2.25}
\end{equation}

We note that $h$ and $b$ are the expectation values of the stress tensor
$T_{\mu\nu}$ and the negative norm tensor $t_{\mu\nu}$ of a log-CFT.  They are related by
\cite{Skenderis:2009nt}
\bea \label{relationsT}
\left\langle {{t_{tt}}} \right\rangle  &= \left\langle {{t_{xx}}} \right\rangle  &=  - \left\langle {{t_{tx}}} \right\rangle  =  - \left\langle {{t_{xt}}} \right\rangle  = \frac{1}
{{4{G_N}}}\{ 2b + h - k\} ,\nonumber \\
\left\langle {{T_{tt}}} \right\rangle  &= \left\langle {{T_{xx}}} \right\rangle  &= \frac{1}
{{4{G_N}}}\{  - 2b + h + k\} ,\nonumber \\
\left\langle {{T_{tx}}} \right\rangle  &= \left\langle {{T_{xt}}} \right\rangle  &= \frac{1}
{{4{G_N}}}\{ 2b + k + h\},
\eea

 Substituting these relations into (\ref{dE})
we have\be
 \begin{split}
  \delta E_{TMG} =&  - \frac{{4\pi }}
{R}\int_{z\to 0, |x|<R, t=t_0} {{\text{d}}x} \{ {h_{tx}} - \rho {\partial _\rho }{h_{tx}} + (x - {x_0})({\partial _t}{h_{xx}} - {\partial _x}{h_{tx}}) \hfill \\
   &- [{R^2} - {(x - {x_0})^2}]{\partial _\rho }{h_{tx}} - 2\rho [{R^2} - {(x - {x_0})^2}]\partial _\rho ^2{h_{tx}}\}  \hfill \\
\end{split}
\label{dE}
\ee
This should be combined with the contribution from the Einstein term, 
\be
\delta E_{EH} = \frac{2\pi }{R}
\int_B {{{{\text{d}}x}}
} \{ [{R^2} - {(x - {x_0})^2}]{\partial _\rho }{h_{tt}} + \frac{1}
{2}({h_{xx}} - {h_{tt}})    + \frac{1}
{{2\rho }}[{R^2} - {(x - {x_0})^2}]({h_{xx}} - {h_{tt}})\},
\ee
which together with the coupling constants, gives an overall energy
\be\label{totalE}
\delta E =2\pi( \frac{1}{16\pi G_N}\delta E_{EH} + \frac{1}{32\pi \mu G_N}\delta E_{TMG}).
\ee
At $\mu=1$, we have
\bea
\delta E&=& \frac{\pi }
{{4{G_N}}}\int_{z\to 0, |x|<R, t=t_0} {\frac{{{\text{d}}x}}
{R}} [{R^2} - {(x - {x_0})^2}]\{ {h_{(2)tt}} + (1 + \log \rho ){b_{(2)tt}}\}  +\nonumber\\
&& \frac{\pi }
{{4{G_N}\mu }}\int_{z\to 0, |x|<R, t=t_0} {\frac{{{\text{d}}x}}
{R}} \{ {b_{(2)tx}}\rho 
   - (x - {x_0})\{ {\partial _t}({h_{(2)xx}}\rho  + {b_{(2)xx}}\rho \log \rho ) -\nonumber\\
 &&{\partial _x}({h_{(2)tx}}\rho  + {b_{(2)tx}}\rho \log \rho )\}  + [{R^2} - {(x - {x_0})^2}][{h_{(2)tx}} + {b_{(2)tx}}(3 + \log \rho )]\}\nonumber\\ &=&  2\pi \int_{-R}^R {{\text{d}}x\frac{{{R^2} - {{(x - {x_0})}^2}}}
{{2R}}} \left\langle {{T_{tt}}} \right\rangle  
\eea
where we have made use of (\ref{relationsT}). The result is precisely the entanglement Hamiltonian (\ref{entangleH}).

\subsubsection{General $\mu$}
It is natural to check also the generic case for $\mu\neq 1$. 
The main difference from the previous case is that the Fefferman-Graham expansion adjusts accordingly:
\be
{h_{ij}} = {h_{( - 2\lambda )ij}}{\rho ^{ - \lambda }} + {h_{(0)ij}} + {h_{(2)ij}}\rho  + {h_{(2 - 2\lambda )ij}}{\rho ^{1 - \lambda }} + {h_{(2 + 2\lambda )ij}}{\rho ^{1 + \lambda }} +  \cdots, \qquad \lambda=\frac{1}{2}(\mu-1).
\ee
We have included here the sources $h_{(-2\lambda)}, h_{0}$ which we will subsequently turn off. We note that the modes sourced by $h_{(-2\lambda)}$ correspond to another operator $X_{ij}$ with conformal dimension $(\Delta_L, \Delta_R)= (2+\lambda,\lambda)$, and they are known to be unstable. The expectation of the stress tensor $T_{ij}$ and $X_{ij}$ are related to the FG coefficients by
\be
\begin{split}
  \left\langle {{T_{ij}}} \right\rangle  =& \frac{1}
{{4{G_N}}}\{ {h_{(2)ij}} - \frac{1}
{{2\lambda  + 1}}{\varepsilon _i}^k{h_{(2)kj}}\}  \hfill \\
  \left\langle {{X_{ij}}} \right\rangle  =& \frac{{\lambda (1 + \lambda )}}
{{4{G_N}(2\lambda  + 1)}}\{ {h_{(2 + 2\lambda )}}_{ij} + {\varepsilon _i}^k{h_{(2 + 2\lambda )}}_{kj}\}  \hfill \\
\end{split}
\ee
Substituting into (\ref{totalE}), we have
\bea
\delta E &=& \frac{\pi }
{{4{G_N}}}\int_{z\to 0, |x|<R, t=t_0} {\frac{{{\text{d}}x}}
{R}} [{R^2} - {(x - {x_0})^2}]\{ {h_{(2)tt}} + (1 + \lambda ){h_{(2 + 2\lambda )tt}}{\rho ^\lambda }\}   \nonumber \\
 &&+ \frac{\pi }
{{4{G_N}\mu }}\int_{z\to 0, |x|<R, t=t_0} {\frac{{{\text{d}}x}}
{R}} \{ \lambda {h_{(2 + 2\lambda )tx}}{\rho ^{\lambda  + 1}} - (x - {x_0})\{ {\partial _t}({h_{(2)xx}}\rho  + {h_{(2 + 2\lambda )xx}}{\rho ^{1 + \lambda }}) \nonumber\\
&&-{\partial _x}({h_{(2)tx}}\rho  + {h_{(2 + 2\lambda )tx}}{\rho ^{1 + \lambda }})\} 
   + [{R^2} - {(x - {x_0})^2}][{h_{(2)tx}} + (2\lambda  + 1)(1 + \lambda ){h_{(2 + 2\lambda )tx}}{\rho ^\lambda }]\}\nonumber\\
&=&2\pi \int_{-R}^R {{\text{d}}x\frac{{{R^2} - {{(x - {x_0})}^2}}}
{{2R}}} \left\langle {{T_{tt}}} \right\rangle
\eea
where the last equality is obtained only assuming that $\lambda>-1, $ corresponding to $\mu>-1$,  since terms proportional to $\rho^{1+\lambda}$ vanish in the boundary $\rho\to 0 $ limit. Reversing the sign of $\mu$ corresponds to exchanging the roles of left and right moving sectors, and we are working here with the assumption that $\mu>0$. The first law of entanglement entropy is thus satisfied. 

\subsection{Entanglement Hamiltonian vs Asymptotic black hole energy  slice}
It is already observed in \cite{Faulkner:2013ica} that the entanglement Hamiltonian at least to linear order in a small perturbation of its expectation value coincides with the asymptotic black hole energy time slice. It is argued that such a first law should follow even for higher derivative gravity so long as the behaviour of the FG expansion near the boundary 
\be
h_{\mu\nu}\sim z^\Delta, 
\ee
 for any $\Delta>2$ (d=2). In the cases we have discussed, we essentially have $\Delta = 2+2\lambda>2$ for any $\lambda>0$. We have checked the extra scenario where $h_{\mu\nu}\sim z^2\log z^2$ and find that this was also sufficient to recover the first law. 

Here we would like to supply one other viewpoint that makes the first law very natural. We recall that the holographic expectation value of the stress tensor of the CFT is obtained by varying the action wrt the boundary metric $h^{(0)}_{\mu\nu}$, the leading zeroth-order term in the FG expansion of $h_{\mu\nu}$:
\be
\langle T^{\mu\nu}\rangle =\frac{-4\pi}{\sqrt{g^{(0)}}} \frac{\delta S}{\delta h_{\mu\nu}^{(0)}}\bigg\vert_{h^{(0)}=0, J_i=0},
\ee 
where $J_i$ denotes a generic source for any other operators. We note that the action $S$ above is given by
\be
S= \int d^{d+1}x \,\, \mathcal{L} + I_{ct},
\ee 
where $I_{ct}$ denotes local counter terms defined at the boundary that removes singularities. 
The above variation is obtained by varying the metric and then integrating by parts to obtain a total derivative term while discarding the equations of motion. Therefore we have
\be
\langle T^{\mu\nu}\rangle \sim \frac{\delta \Theta}{\delta h^{(0)}_{\mu\nu}} + \frac{\delta I_{ct}}{\delta h^{(0)}_{\mu\nu}}
\ee
Now $\Theta$  takes the form   $\Theta \sim \delta g\, \chi,$ so that $\chi$ contributes to $T$. Now on the other hand, in the computation of the asymptotic energy, the  $\Theta$ contribution
to $Q$ came from integrating by part $\delta_\xi g\, \chi$, where $\delta_\xi g $ is given by (\ref{diffeo}). Therefore it gives $\delta E \sim \int \xi.\chi + \cdots$, where $\xi=\xi_B$ and at the boundary reduces precisely to the killing vector of the Rindler horizon preserving the boundary entangling ``surface''.  But both the asymptotic energy and the holographic stress tensor receive extra contributions.  For the asymptotic energy there is contribution from $ \delta_\xi \mathcal{L}$, which can be checked to be proportional to $z^{-1}$ to leading order in the FG expansion, while the expansion of subsequent terms are in powers of $z^2$ and thus do not contribute to further regular terms in the stress tensor. The same can be said of the counter terms in the action. Therefore the variation of the asymptotic energy and the expectation value of the holographic entanglement Hamiltonian should be equal. 

\subsection{First law of a boosted slice }
In the above discussion, we focus on the entanglement entropy of a single interval at a given time slice. There, since it is always possible to identify the bulk entangling surface in pure AdS space with a killing horizon associated to the killing vector $\xi_B$, the holographic entanglement entropy coincides with the black hole entropy, which subsequently guarantees that small perturbations away from the AdS space satisfies a first law relation. The question is, however that if this identification between black hole entropy and holographic entropy is true in one given frame, do we expect this to hold in other inertial frames?

The crucial point is that the holographic entanglement entropy, as noted already in \cite{Castro:2014tta},  coincides with the expression of the black hole entropy when 

\be t^\mu \partial_\mu n =0= t^\mu  \partial_\mu \tilde n, \label{dn}\ee where $n$ and $\tilde n$ are the normals to the bulk entangling surface/horizon and $t$ the tangent. Given such a choice of $n$ and $\tilde n$, the contribution of the TMG term to the (black hole) entropy is given by
\be 
S_{BH}(TMG) =\frac{1}{4G_N\mu} \int_C ds \Gamma^\mu_{\alpha \beta} n^\alpha t^\beta \tilde n_\mu . \label{TMGBH}
\ee
At a killing horizon, the variation gives
\be
\delta S_{BH} = \frac{1}{4G_N\mu}\int_C dx^\beta \delta \Gamma^\mu_{\alpha\beta}\partial_\mu \xi^\alpha
\ee

Generically even if we have fixed a frame such that  $t^\mu\partial_\mu n=0$, this cannot be preserved in a different frame. As a result, the entanglement entropy do not necessarily coincide with the expression (\ref{TMGBH}). The latter however is the quantity that satisfies the first law. Therefore one might worry that when (\ref{dn}) is violated the first law is also violated. The extra terms in the variation of the entanglement entropy is given by
\be
\delta (S_{EE}- S_{BH}) =\frac{1}{4G_N\mu} \int_C \delta g^{\mu\nu} ( \tilde n_\mu\, t^\rho \partial_\rho n_\nu +(\lambda_4n_\mu n_\nu + \lambda_5  \tilde n_\mu \tilde n_\nu)),
\label{disc}\ee
where we have made use of the fact that 
the variation of the entangling surface/bifurcation surface $C$ does not give any contribution, since the curve $C$ is a solution to the equations of motion and the variation leads only to boundary terms which vanishes for fixed boundary condition. Moreover, $\lambda_i$ and $n$ and $\tilde n$ are not varied either since the variation of which also leads to equations of motion. The above terms are the only terms not immediately vanishing up to equations of motion.
The saddle point values of $\lambda_4$ and $\lambda_5$ are given by
\be
2\lambda_4= - n . \nabla\tilde n,\qquad 2\lambda_5=-\tilde n .\nabla n.
\ee
Now substituting into the above expressions:
\be
n= \cosh\eta\, \, q + \sinh \eta \,\,\tilde q\,\qquad \tilde n= \cosh \eta \,\,\tilde q + \sinh\eta\,\,q, 
\ee
where $q, \tilde{q}$ are chosen such that
\be
t^\mu \nabla_\mu q = t^\mu \nabla_\mu \tilde q =0,
\ee
and that  $\eta$ is generically some function of the geodesic coordinate if we were considering some general boosted frame, the three terms in (\ref{disc}) beautifully cancel out.  The holographic entanglement entropy first law is preserved in spite of the difference between the entanglement entropy and the black hole entropy. 
\section{Conclusions }\label{conclude}
In this note, we have computed the variation of the holographic entanglement entropy slightly away from the pure AdS background of a single interval in the presence of the gravitational Chern-Simons term. We found that in a given Lorentz frame the holographic entanglement entropy can still be identified with the entropy associated to a killing horizon, and as such the black hole entropy first law applicable implies a first law of entanglement entropy, subjected to that the perturbation satisfies the linearized equations of motion, regardless of the appearance of unstable modes for generic values of the Chern-Simons coupling $\mu$. However, in a generic Lorentz frame, the gravitational anomaly leads to a discrepancy between the entropy associated to the killing horizon and the holographic entanglement entropy. The variation of these entropies however continues to agree. As a result, the entanglement entropy first law remains satisfied. This is perhaps expected given that the anomaly is a UV property that takes the same value between the excited state and the ground state, which thus cancels out in the difference of the entanglement entropy.
One logical possibility is that effects of non-unitarity can only show up at second order perturbation. We leave these questions for future investigation. 

\section*{Acknowledgements}

We thank FL Lin for reading our manuscript and giving useful comments.
LYH acknowledges support by the Thousand Young Talents Program, and Fudan University.


\appendix
\section{Some useful expressions in deriving TMG Noether charges}
We collect here a set of useful expressions in deriving the TMG Noether charges:
\subsection{$\delta_\xi{\Gamma}$}
\begin{equation}
{{\delta }}\Gamma _\nu ^\kappa  = \frac{1}
{2}{g^{\kappa \lambda }}{{(\delta }}{g_{\mu \lambda ;\nu }} + {{\delta }}{g_{\nu \lambda ;\mu }} - {{\delta }}{g_{\mu \nu ;\lambda }}){\text{d}}{x^\mu }
\label{1.1}
\end{equation}
\begin{equation}
{\delta _\xi }\Gamma _\nu ^\kappa  = ({R^\kappa }_{\nu \rho \mu }{\xi ^\rho } + {\xi ^\kappa }_{;\nu ;\mu }){\text{d}}{x^\mu }
\label{1.2}
\end{equation}
\begin{equation}
{\xi ^\kappa }_{;\nu ;\mu }{\text{d}}{x^\mu } = {\text{d}}{\xi ^\kappa }_{;\nu } + \Gamma _\rho ^\kappa {\xi ^\rho }_{;\nu } - \Gamma _\nu ^\rho {\xi ^\kappa }_{;\rho } = {\text{d}}v_\nu ^\kappa  + \left[ {\Gamma ,v} \right]_\nu ^\kappa
\label{1.3}
\end{equation}
\[v_\nu ^\kappa  = {\xi ^\kappa }_{;\nu }\]
\begin{equation}
{i_\xi }{R^\kappa }_\nu  = {R^\kappa }_{\nu \rho \mu }{\xi ^\rho }{\text{d}}{x^\mu }
\label{1.4}
\end{equation}
\begin{equation}
{\delta _\xi }\Gamma _\nu ^\kappa  = {i_\xi }{R^\kappa }_\nu  + {\text{d}}v_\nu ^\kappa  + \left[ {\Gamma ,v} \right]_\nu ^\kappa
\label{1.5}
\end{equation}
\begin{equation}
{\delta _\xi }{\mathbf{\Gamma }} = {i_\xi }{\mathbf{R}} + {\text{d}}{\mathbf{v}} + \left[ {{\mathbf{\Gamma }},{\mathbf{v}}} \right]
\label{1.6}
\end{equation}

\subsection{$\delta_\xi{R}$}
\begin{equation}
{\mathbf{R}} = {\text{d}}{\mathbf{\Gamma }} + {\mathbf{\Gamma }} \wedge {\mathbf{\Gamma }}
\label{1.7}
\end{equation}
\begin{equation}
\begin{split}
  {\delta _\xi }{\mathbf{R}} =& {\delta _\xi }({\text{d}}{\mathbf{\Gamma }} + {\mathbf{\Gamma }} \wedge {\mathbf{\Gamma }}) \hfill \\
   =& {\text{d}}{i_\xi }{\mathbf{R}} + d\left[ {{\mathbf{\Gamma }},{\mathbf{v}}} \right] + {i_\xi }{\mathbf{R}} \wedge {\mathbf{\Gamma }} + {\text{d}}{\mathbf{v}} \wedge {\mathbf{\Gamma }} + \left[ {{\mathbf{\Gamma }}{\text{,}}{\mathbf{v}}} \right] \wedge {\mathbf{\Gamma }} + {\mathbf{\Gamma }} \wedge {i_\xi }{\mathbf{R}} + {\mathbf{\Gamma }} \wedge {\text{d}}{\mathbf{v}} + {\mathbf{\Gamma }} \wedge \left[ {{\mathbf{\Gamma }},{\mathbf{v}}} \right] \hfill \\
   =& {\text{d}}{i_\xi }{\mathbf{R}} + {\text{d}}{\mathbf{\Gamma v}} - {\mathbf{v}}{\text{d}}{\mathbf{\Gamma }} + {i_\xi }{\mathbf{R}} \wedge {\mathbf{\Gamma }} + {\mathbf{\Gamma }} \wedge {i_\xi }{\mathbf{R}} + \left[ {{\mathbf{\Gamma }} \wedge {\mathbf{\Gamma }}{\text{,}}{\mathbf{v}}} \right] \hfill \\
   =& {\text{d}}{i_\xi }{\mathbf{R}} + \left[ {{\mathbf{R}}{\text{,}}{\mathbf{v}}} \right] + \left\{ {{i_\xi }{\mathbf{R}},{\mathbf{\Gamma }}} \right\} \hfill \\
\end{split}
\label{1.8}
\end{equation}

\subsection{$\delta_\xi L_{TMG}$}
\begin{equation}
L_{TMG} = {\text{tr}}\{ {\mathbf{\Gamma }} \wedge {\mathbf{R}} - \frac{1}
{3}{\mathbf{\Gamma }} \wedge {\mathbf{\Gamma }} \wedge {\mathbf{\Gamma }}\}
\label{1.10}
\end{equation}
\begin{equation}
\begin{split}
  {\delta _\xi }L_{TMG} =& {\delta _\xi }{\text{tr}}\{ {\mathbf{\Gamma }} \wedge {\mathbf{R}} - \frac{1}
{3}{\mathbf{\Gamma }} \wedge {\mathbf{\Gamma }} \wedge {\mathbf{\Gamma }}\}  \hfill \\
   =& {\text{tr}}\{ {\delta _\xi }{\mathbf{\Gamma }} \wedge {\mathbf{R}} + {\mathbf{\Gamma }} \wedge {\delta _\xi }{\mathbf{R}} - {\delta _\xi }{\mathbf{\Gamma }} \wedge {\mathbf{\Gamma }} \wedge {\mathbf{\Gamma }}\}  \hfill \\
   =& {\text{tr}}({i_\xi }{\mathbf{R}} \wedge {\mathbf{R}} + {\text{d}}{\mathbf{v}} \wedge {\mathbf{R}} + \left[ {{\mathbf{\Gamma }},{\mathbf{v}}} \right] \wedge {\mathbf{R}} + {\mathbf{\Gamma }} \wedge {\text{d}}{i_\xi }{\mathbf{R}} + {\mathbf{\Gamma }} \wedge \left[ {{\mathbf{R}},{\mathbf{v}}} \right] + {\mathbf{\Gamma }} \wedge \left\{ {{i_\xi }{\mathbf{R}},{\mathbf{\Gamma }}} \right\} \hfill \\
   &- {i_\xi }{\mathbf{R}} \wedge {\mathbf{\Gamma }} \wedge {\mathbf{\Gamma }} - {\text{d}}{\mathbf{v}} \wedge {\mathbf{\Gamma }} \wedge {\mathbf{\Gamma }} - \left[ {{\mathbf{\Gamma }},{\mathbf{v}}} \right] \wedge {\mathbf{\Gamma }} \wedge {\mathbf{\Gamma }}) \hfill \\
   =& {\text{tr}}({i_\xi }{\mathbf{R}} \wedge {\text{d}}{\mathbf{\Gamma }} + {\text{d}}{\mathbf{v}} \wedge {\text{d}}{\mathbf{\Gamma }} + {\mathbf{\Gamma }} \wedge {\text{d}}{i_\xi }{\mathbf{R}} + {\mathbf{\Gamma }} \wedge \left\{ {{i_\xi }{\mathbf{R}},{\mathbf{\Gamma }}} \right\}) \hfill \\
   =& {\text{tr}}[{\text{d(}}{\mathbf{v}}{\text{d}}{\mathbf{\Gamma }}) - {i_\xi }{\mathbf{R}} \wedge {\text{d}}{\mathbf{\Gamma }} + {\mathbf{\Gamma }} \wedge {\text{d}}{i_\xi }{\mathbf{R}} + 2{i_\xi }{\mathbf{R}} \wedge {\mathbf{R}}] \hfill \\
   =& {\text{tr}}[{\text{d(}}{\mathbf{v}}{\text{d}}{\mathbf{\Gamma }}) - {\text{d}}({\mathbf{\Gamma }} \wedge {i_\xi }{\mathbf{R}}) + 2{i_\xi }{\mathbf{R}} \wedge {\mathbf{R}}] \hfill \\
\end{split}
\label{1.11}
\end{equation}

\subsection{$\Xi_\xi(TMG)$}
\[{\text{d}}{\Xi _\xi(TMG) } \equiv  ({\delta _\xi } - {\mathcal{L}_\xi })L_{TMG}\]
\begin{equation}
\begin{split}
  {\text{d}}{\Xi _\xi(TMG) } =& ({\delta _\xi } - {L_\xi })L_{TMG} \hfill \\
   =& {\text{tr}}\left[ {{\text{d}}({\mathbf{v}}{\text{d}}{\mathbf{\Gamma }}) - {\text{d}}({\mathbf{\Gamma }} \wedge {i_\xi }{\mathbf{R}})} \right] - {\text{d}}({i_\xi }L_{TMG}) \hfill \\
   =& {\text{d}}\{{\text{tr}}[{\mathbf{v}}{\text{d}}{\mathbf{\Gamma }} - {\mathbf{\Gamma }} \wedge {i_\xi }{\mathbf{R}} - ({i_\xi }{\mathbf{\Gamma }} \wedge {\mathbf{R}} - {\mathbf{\Gamma }} \wedge {i_\xi }{\mathbf{R}} - {i_\xi }{\mathbf{\Gamma }} \wedge {\mathbf{\Gamma }} \wedge {\mathbf{\Gamma }})]{\text{\} }} \hfill \\
   =& {\text{d}}\{{\text{tr}}[{\mathbf{v}}{\text{d}}{\mathbf{\Gamma }} - {i_\xi }{\mathbf{\Gamma }} \wedge {\text{d}}{\mathbf{\Gamma }}]{\text{\} }} \hfill \\
   =& {\text{d}}\{ {\text{tr}}[{\mathbf{U}}{\text{d}}{\mathbf{\Gamma }}{\text{]\}  = d}}\{  - {\text{tr}}[{\text{d}}{\mathbf{U}} \wedge {\mathbf{\Gamma }}]\}  \hfill \\
\end{split}
\label{1.14}
\end{equation}
\[\mathbf{U}_\nu ^\kappa  = {\xi ^\kappa }_{,\nu }\]

\subsection{${\Theta_{TMG}}$}
\begin{equation}
\begin{split}
  \delta L_{TMG} =& {\text{tr}}(\delta {\mathbf{\Gamma }} \wedge {\mathbf{R}} + {\mathbf{\Gamma }} \wedge \delta {\mathbf{R}} - \delta {\mathbf{\Gamma }} \wedge {\mathbf{\Gamma }} \wedge {\mathbf{\Gamma }}) \hfill \\
   =& {\text{tr}}[\delta {\mathbf{\Gamma }} \wedge {\mathbf{R}} + \delta {\mathbf{\Gamma }} \wedge {\text{d}}{\mathbf{\Gamma }} + \delta {\mathbf{\Gamma }} \wedge {\mathbf{\Gamma }} \wedge {\mathbf{\Gamma }} - {\text{d}}({\mathbf{\Gamma }} \wedge \delta {\mathbf{\Gamma }})] \hfill \\
   =& {\text{tr}}[2\delta {\mathbf{\Gamma }} \wedge {\mathbf{R}} - {\text{d}}({\mathbf{\Gamma }} \wedge \delta {\mathbf{\Gamma }})] \hfill \\
   =&  - {\text{d}}[{\text{tr}}({\mathbf{\Gamma }} \wedge \delta {\mathbf{\Gamma }})] + \frac{1}
{2}\{ (\delta {g_{\mu \lambda ;\nu }} + \delta {g_{\nu \lambda ;\mu }} - \delta {g_{\mu \nu ;\lambda }}){\text{d}}{x^\mu } \wedge {R^{\nu \lambda }}_{\rho \sigma }{\text{d}}{x^\rho } \wedge {\text{d}}{x^\sigma }\}  \hfill \\
   =&  - {\text{d}}[{\text{tr}}({\mathbf{\Gamma }} \wedge \delta {\mathbf{\Gamma }})] + \frac{1}
{2}\{ {\nabla _\nu }(\delta {g_{\mu \lambda }}{R^{\nu \lambda }}_{\rho \sigma }) + {\nabla _\mu }(\delta {g_{\nu \lambda }}{R^{\nu \lambda }}_{\rho \sigma }) - {\nabla _\lambda }(\delta {g_{\mu \nu }}{R^{\nu \lambda }}_{\rho \sigma })\} {\text{d}}{x^\mu } \wedge {\text{d}}{x^\rho } \wedge {\text{d}}{x^\sigma } \hfill \\
   &- \frac{1}
{2}\{ \delta {g_{\mu \lambda }}{\nabla _\nu }{R^{\nu \lambda }}_{\rho \sigma } + \delta {g_{\nu \lambda }}{\nabla _\mu }{R^{\nu \lambda }}_{\rho \sigma } - \delta {g_{\mu \nu }}{\nabla _\lambda }{R^{\nu \lambda }}_{\rho \sigma }\} {\text{d}}{x^\mu } \wedge {\text{d}}{x^\rho } \wedge {\text{d}}{x^\sigma } \hfill \\
   =&  - {\text{d}}[{\text{tr}}({\mathbf{\Gamma }} \wedge \delta {\mathbf{\Gamma }})] + \{ {\nabla _\nu }(\delta {g_{\mu \lambda }}{R^{\nu \lambda }}_{\rho \sigma }) + \delta {g_{\mu \nu }}{\nabla _\lambda }{R^{\nu \lambda }}_{\rho \sigma }\} {\text{d}}{x^\mu } \wedge {\text{d}}{x^\rho } \wedge {\text{d}}{x^\sigma } \hfill \\
   =&  - {\text{d}}[{\text{tr}}({\mathbf{\Gamma }} \wedge \delta {\mathbf{\Gamma }})] + {\partial _\nu }(\delta {g_{\mu \lambda }}{R^{\nu \lambda }}_{\rho \sigma }{{\hat \epsilon }^{\mu \rho \sigma }}){\text{d}}{x^1} \wedge {\text{d}}{x^2} \wedge {\text{d}}{x^3} \hfill \\
   &+ \delta {g_{\mu \nu }}{\nabla _\lambda }{R^{\nu \lambda }}_{\rho \sigma }{{\hat \epsilon }^{\mu \rho \sigma }}{\text{d}}{x^1} \wedge {\text{d}}{x^2} \wedge {\text{d}}{x^3} \hfill \\
   =&  - {\text{d}}[{\text{tr}}({\mathbf{\Gamma }} \wedge \delta {\mathbf{\Gamma }})] + {\partial _\nu }(\delta {g_{\mu \lambda }}{R^{\nu \lambda }}_{\rho \sigma }{{\hat \epsilon }^{\mu \rho \sigma }}){\text{d}}{x^1} \wedge {\text{d}}{x^2} \wedge {\text{d}}{x^3} \hfill \\
   &+ \frac{1}
{2}\delta {g_{\mu \nu }}\{ {\nabla _\lambda }{R^{\nu \lambda }}_{\rho \sigma }{{\hat \epsilon }^{\mu \rho \sigma }} + {\nabla _\lambda }{R^{\mu \lambda }}_{\rho \sigma }{{\hat \epsilon }^{\nu \rho \sigma }}\} {\text{d}}{x^1} \wedge {\text{d}}{x^2} \wedge {\text{d}}{x^3} \hfill \\
   =&  - {\text{d}}[{\text{tr}}({\mathbf{\Gamma }} \wedge \delta {\mathbf{\Gamma }})] + {\partial _\nu }(\delta {g_{\mu \lambda }}{R^{\nu \lambda }}_{\rho \sigma }{{\hat \epsilon }^{\mu \rho \sigma }}){\text{d}}{x^1} \wedge {\text{d}}{x^2} \wedge {\text{d}}{x^3} + ({\text{EoM}}{\text{.}}) \hfill \\
\end{split}
\label{1.16}
\end{equation}

\subsection{${\Pi _\xi(TMG) }$}
\[{\Pi _\xi(TMG) } \equiv ({\delta _\xi } - {\mathcal{L}_\xi })\Theta_{TMG} \]
\begin{equation}
{\Pi _\xi(TMG) } = ({\delta _\xi } - {\mathcal{L}_\xi })\Theta_{TMG}  =  - ({\delta _\xi } - {\mathcal{L}_\xi }){\text{tr}}({\mathbf{\Gamma }} \wedge \delta {\mathbf{\Gamma }}) =  - {\text{tr}}[({\delta _\xi } - {\mathcal{L}_\xi }){\mathbf{\Gamma }} \wedge \delta {\mathbf{\Gamma }}]
\label{1.18}
\end{equation}
\begin{equation}
\begin{split}
  {\mathcal{L}_\xi }\Gamma _\nu ^\kappa  =& {\xi ^\rho }{\partial _\rho }\Gamma _\nu ^\kappa  - \Gamma _\nu ^\rho {\xi ^\kappa }_{,\rho } + \Gamma _\rho ^\kappa {\xi ^\rho }_{,\nu } + \Gamma _{\rho \nu }^\kappa {\xi ^\rho }_{,\mu }{\text{d}}{x^\mu } \hfill \\
   =& {\xi ^\rho }{\partial _\rho }\Gamma _\nu ^\kappa  + \Gamma _{\rho \nu }^\kappa {\xi ^\rho }_{,\mu }{\text{d}}{x^\mu } - \Gamma _\nu ^\rho {\xi ^\kappa }_{;\rho } + \Gamma _\nu ^\rho \Gamma _{\rho \sigma }^\kappa {\xi ^\sigma } + \Gamma _\rho ^\kappa {\xi ^\rho }_{;\nu } - \Gamma _\rho ^\kappa \Gamma _{\nu \sigma }^\rho {\xi ^\sigma } \hfill \\
\end{split}
\label{1.19}
\end{equation}
\begin{equation}
{\mathcal{L}_\xi }{\mathbf{\Gamma }} = {\xi ^\rho }{\partial _\rho }{\mathbf{\Gamma }} + {{\mathbf{\Gamma }}_\rho }{\xi ^\rho }_{,\mu }{\text{d}}{x^\mu } + \left[ {{\mathbf{\Gamma }},{\mathbf{v}}} \right] + {i_\xi }({\mathbf{\Gamma }} \wedge {\mathbf{\Gamma }})
\label{1.20}
\end{equation}
\begin{equation}
\begin{split}
  ({\delta _\xi } - {\mathcal{L}_\xi })\Gamma  =& {i_\xi }{\mathbf{R}} + {\text{d}}{\mathbf{v}} - {\xi ^\rho }{\partial _\rho }{\mathbf{\Gamma }} - {{\mathbf{\Gamma }}_\rho }{\xi ^\rho }_{,\mu }{\text{d}}{x^\mu } - {i_\xi }({\mathbf{\Gamma }} \wedge {\mathbf{\Gamma }}) \hfill \\
   =& {i_\xi }{\text{d}}{\mathbf{\Gamma }} + {\text{d}}{\mathbf{U}} - {\text{d}}{i_\xi }{\mathbf{\Gamma }} - {{\mathbf{\Gamma }}_\rho }{\text{d}}{\xi ^\rho } \hfill \\
   =& {\text{d}}{\mathbf{U}} + {i_\xi }{\text{d}}{\mathbf{\Gamma }} - {i_\xi }{\text{d}}{\mathbf{\Gamma }} = {\text{d}}{\mathbf{U}} \hfill \\
\end{split}
\label{1.21}
\end{equation}
\begin{equation}
{\Pi _\xi } (TMG)=  - {\text{tr}}[({\delta _\xi } - {\mathcal{L}_\xi }){\mathbf{\Gamma }} \wedge \delta {\mathbf{\Gamma }}] =  - tr\left[ {{\text{d}}{\mathbf{U}} \wedge \delta {\mathbf{\Gamma }}} \right]
\label{1.22}
\end{equation}

\subsection{$\Sigma_{\xi}(TMG)$}
\begin{equation}
\text{d}{\Sigma _\xi(TMG) } = {\Pi _\xi(TMG) } - \delta {\Xi _\xi(TMG) } = 0
\label{1.23}
\end{equation}
Throughout this paper, our choice of boundary terms correspond to \cite{Tachikawa:2006sz}:
\begin{equation}
{\Sigma _\xi(TMG) } = 0.
\label{1.24}
\end{equation}

\subsection{$J_\xi(TMG)$}
\[{J_\xi(TMG) } = {\Theta _\xi(TMG) } - {i_\xi }L(TMG) - {\Xi _\xi(TMG) }\]
\begin{equation}
\begin{split}
  {J_\xi(TMG) } =& {\Theta _\xi(TMG) } - {i_\xi }L(TMG) - {\Xi _\xi(TMG) } \hfill \\
   =&  - {\text{tr}}({\mathbf{\Gamma }} \wedge {\delta _\xi }{\mathbf{\Gamma }}) + \frac{1}
{2}\delta_\xi {g_{\mu \lambda }}{R^{\nu \lambda }}_{\rho \sigma }{{\hat \epsilon }^{\mu \rho \sigma }}{{\hat \epsilon }_{\nu \alpha \beta }}{\text{d}}{x^\alpha } \wedge {\text{d}}{x^\beta } \hfill \\
   &- {\text{tr}}({i_\xi }{\mathbf{\Gamma }} \wedge {\mathbf{R}} - {\mathbf{\Gamma }} \wedge {i_\xi }{\mathbf{R}} - {i_\xi }{\mathbf{\Gamma }} \wedge {\mathbf{\Gamma }} \wedge {\mathbf{\Gamma }}) + {\text{tr}}[{\text{d}}{\mathbf{U}} \wedge {\mathbf{\Gamma }}] \hfill \\
   =& \frac{1}
{2}({\nabla _\mu }{\xi _\lambda } + {\nabla _\lambda }{\xi _\mu })
{R^{\nu \lambda }}_{\rho \sigma }{{\hat \epsilon }^{\mu \rho \sigma }}{{\hat \epsilon }_{\nu \alpha \beta }}{\text{d}}{x^\alpha } \wedge {\text{d}}{x^\beta } \hfill \\
   &- {\text{tr}}({\mathbf{\Gamma }} \wedge {\text{d}}{\mathbf{v}} + 2{\mathbf{\Gamma }} \wedge {\mathbf{\Gamma v}} + {i_\xi }{\mathbf{\Gamma }} \wedge {\text{d}}{\mathbf{\Gamma }} - {\text{d}}{\mathbf{U}} \wedge {\mathbf{\Gamma }}) \hfill \\
   =& \frac{1}
{2}({\nabla _\mu }{\xi _\lambda } + {\nabla _\lambda }{\xi _\mu })
{R^{\nu \lambda }}_{\rho \sigma }{{\hat \epsilon }^{\mu \rho \sigma }}{{\hat \epsilon }_{\nu \alpha \beta }}{\text{d}}{x^\alpha } \wedge {\text{d}}{x^\beta } \hfill \\
   &- {\text{tr}}[2{\mathbf{Rv}} - 2{\text{d}}{\mathbf{\Gamma v}} + {\mathbf{\Gamma }} \wedge {\text{d}}{\mathbf{v}} + {\text{d}}({i_\xi }{\mathbf{\Gamma }} \wedge {\mathbf{\Gamma }}) - {\text{d}}{i_\xi }{\mathbf{\Gamma }} \wedge {\mathbf{\Gamma }} - {\text{d}}{\mathbf{U}} \wedge {\mathbf{\Gamma }}] \hfill \\
   =& \frac{1}
{2}({\nabla _\mu }{\xi _\lambda } + {\nabla _\lambda }{\xi _\mu })
{R^{\nu \lambda }}_{\rho \sigma }{{\hat \epsilon }^{\mu \rho \sigma }}{{\hat \epsilon }_{\nu \alpha \beta }}{\text{d}}{x^\alpha } \wedge {\text{d}}{x^\beta } - {\text{tr}}[2{\mathbf{Rv}} - 2{\text{d}}{\mathbf{\Gamma v}} + 2{\mathbf{\Gamma }} \wedge {\text{d}}{\mathbf{v}} + {\text{d}}({i_\xi }{\mathbf{\Gamma }} \wedge {\mathbf{\Gamma }})] \hfill \\
   =& \frac{1}
{2}({\nabla _\mu }{\xi _\lambda } + {\nabla _\lambda }{\xi _\mu })
{R^{\nu \lambda }}_{\rho \sigma }{{\hat \epsilon }^{\mu \rho \sigma }}{{\hat \epsilon }_{\nu \alpha \beta }}{\text{d}}{x^\alpha } \wedge {\text{d}}{x^\beta } - {\text{tr}}(2{\mathbf{Rv}}) + {\text{d}}[{\text{tr}}(2{\mathbf{v\Gamma }} - {i_\xi }{\mathbf{\Gamma }} \wedge {\mathbf{\Gamma }})] \hfill \\
\end{split}
\label{1.25}
\end{equation}

\subsection{$Q_\xi(TMG)$}
\begin{equation}
\begin{split}
  {J_\xi } (TMG)=& \frac{1}
{2}({\nabla _\mu }{\xi _\lambda } + {\nabla _\lambda }{\xi _\mu })
{R^{\nu \lambda }}_{\rho \sigma }{{\hat \epsilon }^{\mu \rho \sigma }}{{\hat \epsilon }_{\nu \alpha \beta }}{\text{d}}{x^\alpha } \wedge {\text{d}}{x^\beta } - {\text{tr}}(2{\mathbf{Rv}}) + {\text{d}}[{\text{tr}}(2{\mathbf{v\Gamma }} - {i_\xi }{\mathbf{\Gamma }} \wedge {\mathbf{\Gamma }})] \hfill \\
   =& {\text{d}}[{\text{tr}}(2{\mathbf{v\Gamma }} - {i_\xi }{\mathbf{\Gamma }} \wedge {\mathbf{\Gamma }})] \hfill \\
   &+ \frac{1}
{2}({\nabla _\mu }{\xi _\lambda } + {\nabla _\lambda }{\xi _\mu }){R^{\nu \lambda }}_{\rho \sigma }{{\hat \epsilon }^{\mu \rho \sigma }}{{\hat \epsilon }_{\nu \alpha \beta }}{\text{d}}{x^\alpha } \wedge {\text{d}}{x^\beta } - \frac{1}
{2}{\nabla _\lambda }{\xi _\mu }{R^{\lambda \mu }}_{\rho \sigma }{{\hat \epsilon }^{\nu \rho \sigma }}{{\hat \epsilon }_{\nu \alpha \beta }}d{x^\alpha } \wedge d{x^\beta } \hfill \\
   =& {\text{d}}[{\text{tr}}(2{\mathbf{v\Gamma }} - {i_\xi }{\mathbf{\Gamma }} \wedge {\mathbf{\Gamma }})] + \frac{1}
{2}{\nabla _\lambda }{\xi _\mu }({R^{\nu \mu }}_{\rho \sigma }{{\hat \epsilon }^{\lambda \rho \sigma }} + {R^{\nu \lambda }}_{\rho \sigma }{{\hat \epsilon }^{\mu \rho \sigma }} - {R^{\lambda \mu }}_{\rho \sigma }{{\hat \epsilon }^{\nu \rho \sigma }}){{\hat \epsilon }_{\nu \alpha \beta }}{\text{d}}{x^\alpha } \wedge {\text{d}}{x^\beta } \hfill \\
   =& {\text{d}}[{\text{tr}}(2{\mathbf{v\Gamma }} - {i_\xi }{\mathbf{\Gamma }} \wedge {\mathbf{\Gamma }})] + \frac{1}
{2}{\nabla _\lambda }[{\xi _\mu }({R^{\nu \mu }}_{\rho \sigma }{{\hat \epsilon }^{\lambda \rho \sigma }} + {R^{\nu \lambda }}_{\rho \sigma }{{\hat \epsilon }^{\mu \rho \sigma }} - {R^{\lambda \mu }}_{\rho \sigma }{{\hat \epsilon }^{\nu \rho \sigma }})]{{\hat \epsilon }_{\nu \alpha \beta }}{\text{d}}{x^\alpha } \wedge {\text{d}}{x^\beta } \hfill \\
   &- \frac{1}
{2}{\xi _\mu }{\nabla _\lambda }({R^{\nu \mu }}_{\rho \sigma }{{\hat \epsilon }^{\lambda \rho \sigma }} + {R^{\nu \lambda }}_{\rho \sigma }{{\hat \epsilon }^{\mu \rho \sigma }} - {R^{\lambda \mu }}_{\rho \sigma }{{\hat \epsilon }^{\nu \rho \sigma }}){{\hat \epsilon }_{\nu \alpha \beta }}{\text{d}}{x^\alpha } \wedge {\text{d}}{x^\beta } \hfill \\
   =& {\text{d}}[{\text{tr}}(2{\mathbf{v\Gamma }} - {i_\xi }{\mathbf{\Gamma }} \wedge {\mathbf{\Gamma }})] + \frac{1}
{2}{\partial _\lambda }[{\xi _\mu }({R^{\nu \mu }}_{\rho \sigma }{{\hat \epsilon }^{\lambda \rho \sigma }} + {R^{\nu \lambda }}_{\rho \sigma }{{\hat \epsilon }^{\mu \rho \sigma }} - {R^{\lambda \mu }}_{\rho \sigma }{{\hat \epsilon }^{\nu \rho \sigma }})]{{\hat \epsilon }_{\nu \alpha \beta }}{\text{d}}{x^\alpha } \wedge {\text{d}}{x^\beta } \hfill \\
   &+ 2{\xi _\mu } \cdot \frac{1}
{2}{\nabla _\lambda }({R^{\lambda \nu }}_{\rho \sigma }{{\hat \epsilon }^{\mu \rho \sigma }} + {R^{\lambda \mu }}_{\rho \sigma }{{\hat \epsilon }^{\nu \rho \sigma }})\frac{1}
{2}{{\hat \epsilon }_{\nu \alpha \beta }}{\text{d}}{x^\alpha } \wedge {\text{d}}{x^\beta } \hfill \\
   =& {\text{d}}[{\text{tr}}(2{\mathbf{v\Gamma }} - {i_\xi }{\mathbf{\Gamma }} \wedge {\mathbf{\Gamma }})] + \frac{1}
{2}{\partial _\lambda }[{\xi _\mu }({R^{\nu \mu }}_{\rho \sigma }{{\hat \epsilon }^{\lambda \rho \sigma }} + {R^{\nu \lambda }}_{\rho \sigma }{{\hat \epsilon }^{\mu \rho \sigma }} - {R^{\lambda \mu }}_{\rho \sigma }{{\hat \epsilon }^{\nu \rho \sigma }})]{{\hat \epsilon }_{\nu \alpha \beta }}{\text{d}}{x^\alpha } \wedge {\text{d}}{x^\beta } \hfill \\
   &+ 2{\xi _\mu }{({\text{EoM}}{\text{.}})^{\mu \nu }}\frac{1}
{2}{{\hat \epsilon }_{\nu \alpha \beta }}{\text{d}}{x^\alpha } \wedge {\text{d}}{x^\beta } \hfill \\
\end{split}
\label{1.27}
\end{equation}

\begin{equation}
{Q_\xi } = {\text{tr}}(2{\mathbf{v\Gamma }} - {i_\xi }{\mathbf{\Gamma }} \wedge {\mathbf{\Gamma }}) + \frac{1}
{2}{\xi _\mu }({R^{\nu \mu }}_{\rho \sigma }{{\hat \epsilon }^{\lambda \rho \sigma }} + {R^{\nu \lambda }}_{\rho \sigma }{{\hat \epsilon }^{\mu \rho \sigma }} - {R^{\lambda \mu }}_{\rho \sigma }{{\hat \epsilon }^{\nu \rho \sigma }}){{\hat \epsilon }_{\nu \lambda \alpha }}{\text{d}}{x^\alpha }
\label{1.28}
\end{equation}

\subsection{${{Q'}_\xi }$}
\[\delta {C_\xi } \equiv {i_\xi }\Theta  + {\Sigma _\xi }\]
\[{{Q'}_\xi } \equiv {Q_\xi } - {C_\xi }\]
\begin{equation}
\begin{split}
  \delta {{Q'}_\xi } \equiv& \delta {Q_\xi } - \delta {C_\xi } \hfill \\
   =& \delta {\text{tr}}(2{\mathbf{U\Gamma }} + {i_\xi }{\mathbf{\Gamma }} \wedge {\mathbf{\Gamma }}) + \frac{1}
{2}\delta \{ {\xi _\mu }({R^{\nu \mu }}_{\rho \sigma }{{\hat \epsilon }^{\lambda \rho \sigma }} + {R^{\nu \lambda }}_{\rho \sigma }{{\hat \epsilon }^{\mu \rho \sigma }} - {R^{\lambda \mu }}_{\rho \sigma }{{\hat \epsilon }^{\nu \rho \sigma }}){{\hat \epsilon }_{\nu \lambda \beta }}{\text{d}}{x^\beta }\}  \hfill \\
   &+ {i_\xi }{\text{tr}}({\mathbf{\Gamma }} \wedge \delta {\mathbf{\Gamma }}) - {\xi ^\alpha }\delta {g_{\mu \lambda }}{R^{\nu \lambda }}_{\rho \sigma }{{\hat \epsilon }^{\mu \rho \sigma }}{{\hat \epsilon }_{\nu \alpha \beta }}{\text{d}}{x^\beta } \hfill \\
   =& {\text{tr}}(2{\mathbf{U}}\delta {\mathbf{\Gamma }} + 2{i_\xi }{\mathbf{\Gamma }} \wedge \delta {\mathbf{\Gamma }}) + \frac{1}
{2}\delta \{ {\xi _\mu }( - 4{R^\mu }_\beta  + {R^{\nu \lambda }}_{\rho \sigma }{{\hat \epsilon }^{\mu \rho \sigma }}{{\hat \epsilon }_{\nu \lambda \beta }})\} {\text{d}}{x^\beta } \hfill \\
   &- {\xi ^\alpha }\delta {g_{\mu \lambda }}{R^{\nu \lambda }}_{\rho \sigma }{{\hat \epsilon }^{\mu \rho \sigma }}{{\hat \epsilon }_{\nu \alpha \beta }}{\text{d}}{x^\beta } \hfill \\
\end{split}
\label{1.29}
\end{equation}
\begin{equation}
{R^{\nu \lambda }}_{\rho \sigma } = (\delta _\rho ^\nu {R^\lambda }_\sigma  - \delta _\sigma ^\nu {R^\lambda }_\rho  - \delta _\rho ^\lambda {R^\nu }_\sigma  + \delta _\sigma ^\lambda {R^\nu }_\rho ) - \frac{1}
{2}R(\delta _\rho ^\nu \delta _\sigma ^\lambda  - \delta _\rho ^\lambda \delta _\sigma ^\nu )
\label{1.30}
\end{equation}
where we recall that $U^\mu_\nu= \partial_\nu \xi^\mu$.
\begin{equation}
\begin{split}
  \delta {{Q'}_\xi } =& {\text{tr}}(2{\mathbf{U}}\delta {\mathbf{\Gamma }} + 2{i_\xi }{\mathbf{\Gamma }} \wedge \delta {\mathbf{\Gamma }}) + \frac{1}
{2}\delta \{ {\xi _\mu }( - 4{R^\mu }_\beta  + 2R\delta _\rho ^\mu  - 4{R^\mu }_\beta )\} {\text{d}}{x^\beta } \hfill \\
  &- 2\delta {g_{\mu \lambda }}({\xi ^\alpha }  {R^\lambda }_\alpha {\text{d}}{x^\mu } - {\xi ^\mu }  {R^\lambda }_\alpha {\text{d}}{x^\alpha }) \hfill \\
   =& {\text{tr}}(2{\mathbf{U}}\delta {\mathbf{\Gamma }} + 2{i_\xi }{\mathbf{\Gamma }} \wedge \delta {\mathbf{\Gamma }}) + \delta \{ {\xi _\mu }(R\delta _\beta ^\mu  - 4{R^\mu }_\beta )\} {\text{d}}{x^\beta } -2 \delta {g_{\mu \lambda }}({\xi ^\alpha }{R^\lambda }_\alpha {\text{d}}{x^\mu } - {\xi ^\mu }{R^\lambda }_\alpha {\text{d}}{x^\alpha }) \hfill \\
   =& {\text{tr}}(2{\mathbf{U}}\delta {\mathbf{\Gamma }} + 2{i_\xi }{\mathbf{\Gamma }} \wedge \delta {\mathbf{\Gamma }}) + \delta \{ {\xi ^\alpha }({g_{\alpha \beta }}R - 4{R_{\alpha \beta }})\} {\text{d}}{x^\beta } -2 \delta {g_{\mu \lambda }}({\xi ^\alpha }{R^\lambda }_\alpha {\text{d}}{x^\mu } - {\xi ^\mu }{R^\lambda }_\alpha {\text{d}}{x^\alpha }) \hfill \\
   =& {\text{tr}}(2{\mathbf{U}}\delta {\mathbf{\Gamma }} + 2{i_\xi }{\mathbf{\Gamma }} \wedge \delta {\mathbf{\Gamma }}) + {\xi ^\alpha }(\delta {g_{\alpha \beta }}R + {g_{\alpha \beta }}\delta R - 4\delta {R_{\alpha \beta }}){\text{d}}{x^\beta } \hfill \\
  & -2 \delta {g_{\mu \lambda }}{g^{\lambda \beta }}({\xi ^\alpha }{R_{\beta \alpha }}{\text{d}}{x^\mu } - {\xi ^\mu }{R_{\beta \alpha }}{\text{d}}{x^\alpha }) \hfill \\
\end{split}
\label{1.31}
\end{equation}
\section{Some useful expressions in TMG holographic dictionary}
This is a collection of useful expressions reading off expectation values of various operators of the dual CFT from the bulk metric. These expressions are collected from \cite{Skenderis:2009nt}.
\subsection{At $\mu=1$}
\begin{equation}
{h_{ij}} = {b_{(0)ij}}\log \rho  + {h_{(0)ij}} + {b_{(2)ij}}\rho \log \rho  + {h_{(2)ij}}\rho  + ...
\label{2.18}
\end{equation}
\begin{equation}
\begin{split}
  \left\langle {{T_{ij}}} \right\rangle  =& \frac{1}
{{4{G_N}}}\{ {b_{(2)ij}} - 3{\varepsilon _i}^k{b_{(2)kj}} + 2\bar P_i^k{h_{(2)kj}} + {\eta _{ij}}(\frac{1}
{2}\tilde R[{h_{(0)ij}}] + \tilde R[{b_{(0)ij}}]) \hfill \\
  & + \frac{1}
{2}\bar P_i^k({\partial ^l}{\partial _l}{h_{(0)kj}} - {\partial _j}{\partial ^l}{h_{(0)lk}})\} , \hfill \\
  \left\langle {{t_{ij}}} \right\rangle  =& \frac{1}
{{2{G_N}}}P_i^k\{ {b_{(2)ij}} + {h_{(2)ij}} - \frac{1}
{2}{\eta _{kj}}{\text{tr}}({b_{(2)}} + {h_{(2)}})\} . \hfill \\
\end{split}
\label{2.19}
\end{equation}
\begin{equation}
{h_{ij}} = {b_{(2)ij}}\rho \log \rho  + {h_{(2)ij}}\rho  + ...
\label{2.20}
\end{equation}
\begin{equation}
\begin{split}
  {\text{tr}}({h_{(2)}}) =& {\text{tr}}({b_{(2)}}) = 0, \hfill \\
  {b_{(2)ij}} - {\varepsilon _i}^k{b_{(2)kj}} = 0&,{\partial ^j}{h_{(2)ji}} - {\varepsilon _i}^k{\partial ^j}{h_{(2)jk}} = 0 \hfill \\
\end{split}
\label{2.21}
\end{equation}
\begin{equation}
\begin{split}
  \left\langle {{T_{ij}}} \right\rangle  =&  \frac{1}
{{4{G_N}}}\{ {b_{(2)ij}} - 3{\varepsilon _i}^k{b_{(2)kj}} + {h_{(2)ij}} - {\varepsilon _i}^k{h_{(2)kj}}\} , \hfill \\
  \left\langle {{t_{ij}}} \right\rangle  =& \frac{1}
{{4{G_N}}}\{ {b_{(2)ij}} + {\varepsilon _i}^k{b_{(2)kj}} + {h_{(2)ij}} + {\varepsilon _i}^k{h_{(2)kj}}\}  \hfill \\
\end{split} 
\label{2.22}
\end{equation}
\begin{equation}
\begin{split}
  \left\langle {{T_{tt}}} \right\rangle  =& \frac{1}
{{4{G_N}}}\{ {b_{(2)tt}} + 3{b_{(2)xt}} + {h_{(2)tt}} + {h_{(2)xt}}\} , \hfill \\
  \left\langle {{T_{tx}}} \right\rangle  =& \frac{1}
{{4{G_N}}}\{ {b_{(2)tx}} + 3{b_{(2)xx}} + {h_{(2)tx}} + {h_{(2)xx}}\} , \hfill \\
  \left\langle {{T_{xt}}} \right\rangle  =& \frac{1}
{{4{G_N}}}\{ {b_{(2)xt}} + 3{b_{(2)tt}} + {h_{(2)xt}} + {h_{(2)tt}}\} , \hfill \\
  \left\langle {{T_{xx}}} \right\rangle  =& \frac{1}
{{4{G_N}}}\{ {b_{(2)xx}} + 3{b_{(2)tx}} + {h_{(2)xx}} + {h_{(2)tx}}\} , \hfill \\
\end{split} 
\label{2.23}
\end{equation}
\begin{equation}
\begin{split}
  \left\langle {{t_{tt}}} \right\rangle  =& \frac{1}
{{4{G_N}}}\{ {b_{(2)tt}} - {b_{(2)xt}} + {h_{(2)tt}} - {h_{(2)xt}}\} , \hfill \\
  \left\langle {{t_{tx}}} \right\rangle  =& \frac{1}
{{4{G_N}}}\{ {b_{(2)tx}} - {b_{(2)xx}} + {h_{(2)tx}} - {h_{(2)xx}}\} , \hfill \\
  \left\langle {{t_{xt}}} \right\rangle  =& \frac{1}
{{4{G_N}}}\{ {b_{(2)xt}} - {b_{(2)tt}} + {h_{(2)xt}} - {h_{(2)tt}}\} , \hfill \\
  \left\langle {{t_{xx}}} \right\rangle  =& \frac{1}
{{4{G_N}}}\{ {b_{(2)xx}} - {b_{(2)tx}} + {h_{(2)xx}} - {h_{(2)tx}}\} . \hfill \\
\end{split}
\label{2.24} 
\end{equation}
\begin{equation}
{\mathbf{h}} = \rho \log \rho \left( {\begin{array}{*{20}{c}}
   b & { - b}  \\
   { - b} & b  \\

 \end{array} } \right) + \rho \left( {\begin{array}{*{20}{c}}
   h & k  \\
   k & h  \\

 \end{array} } \right) + ...
 \label{2.25}
\end{equation}
\begin{equation}
  ({\partial ^x} + {\partial ^t})(h + k) = 0 \text{,   i.e.  }\bar \partial T = 0
  \label{2.26}
\end{equation}
\begin{equation}
\begin{gathered}
  \left\langle {{t_{tt}}} \right\rangle  = \left\langle {{t_{xx}}} \right\rangle  =  - \left\langle {{t_{tx}}} \right\rangle  =  - \left\langle {{t_{xt}}} \right\rangle  = \frac{1}
{{4{G_N}}}\{ 2b + h - k\} , \hfill \\
  \left\langle {{T_{tt}}} \right\rangle  = \left\langle {{T_{xx}}} \right\rangle  = \frac{1}
{{4{G_N}}}\{  - 2b + h + k\} , \hfill \\
  \left\langle {{T_{tx}}} \right\rangle  = \left\langle {{T_{xt}}} \right\rangle  = \frac{1}
{{4{G_N}}}\{ 2b + k + h\} , \hfill \\
\end{gathered}
\label{2.27}
\end{equation}

\subsection{At $\mu \neq 1$}

\begin{equation}
{h_{ij}} = {h_{( - 2\lambda )ij}}{\rho ^{ - \lambda }} + {h_{(0)ij}} + {h_{(2)ij}}\rho  + {h_{(2 - 2\lambda )ij}}{\rho ^{1 - \lambda }} + {h_{(2 + 2\lambda )ij}}{\rho ^{1 + \lambda }} +  \cdots \text{,  } \lambda=\frac{1}{2}(\mu-1)
\label{3.8}
\end{equation}
\[h_{(0)ij}=0\text{,   }{h_{( - 2\lambda )ij}}=0\]
\begin{equation}
\begin{split}
  {\text{tr}}({h_{( - 2\lambda )}}) = {\text{tr}}({h_{(2)}}) = {\text{tr}}({h_{(2 - 2\lambda )}}) = {\text{tr}}({h_{(2 + 2\lambda )}}) = 0 \hfill \\
  {h_{(2 + 2\lambda )ij}} = {\varepsilon _i}^k{h_{(2 + 2\lambda )kj}},{h_{(2 - 2\lambda )ij}} = 0 \hfill \\
\end{split}
\label{3.9}
\end{equation}
\begin{equation}
\begin{split}
  \left\langle {{T_{ij}}} \right\rangle  = &\frac{1}
{{4{G_N}}}\{ (\delta _i^k - \frac{1}
{{2\lambda  + 1}}{\varepsilon _i}^k){h_{(2)kj}} - {\eta _{ij}}{\text{tr}}({h_{(2)}}) + \frac{1}
{{2(2\lambda  + 1)}}\bar P_i^k({\partial ^l}{\partial _l}{h_{(0)kj}} - {\partial _j}{\partial ^l}{h_{(0)kl}})\}  \hfill \\
  \left\langle {{X_{ij}}} \right\rangle  =& \frac{{\lambda (1 + \lambda )}}
{{2{G_N}(2\lambda  + 1)}}{({h_{(2 + 2\lambda )}}_{ij})_L} \hfill \\
\end{split}
\label{3.10}
\end{equation}
\begin{equation}
\begin{split}
  \left\langle {{T_{ij}}} \right\rangle  =& \frac{1}
{{4{G_N}}}\{ {h_{(2)ij}} - \frac{1}
{{2\lambda  + 1}}{\varepsilon _i}^k{h_{(2)kj}}\}  \hfill \\
  \left\langle {{X_{ij}}} \right\rangle  =& \frac{{\lambda (1 + \lambda )}}
{{4{G_N}(2\lambda  + 1)}}\{ {h_{(2 + 2\lambda )}}_{ij} + {\varepsilon _i}^k{h_{(2 + 2\lambda )}}_{kj}\}  \hfill \\
\end{split}
\label{3.11}
\end{equation}
\begin{equation}
\begin{split}
  \left\langle {{T_{tt}}} \right\rangle  =& \frac{1}
{{4{G_N}}}\{ {h_{(2)tt}} + \frac{1}
{{2\lambda  + 1}}{h_{(2)xt}}\}    \hfill \\
  \left\langle {{T_{tx}}} \right\rangle  = &\frac{1}
{{4{G_N}}}\{ {h_{(2)tx}} + \frac{1}
{{2\lambda  + 1}}{h_{(2)xx}}\}   \hfill \\
  \left\langle {{T_{xt}}} \right\rangle  =& \frac{1}
{{4{G_N}}}\{ {h_{(2)xt}} + \frac{1}
{{2\lambda  + 1}}{h_{(2)tt}}\} = \left\langle {{T_{tx}}} \right\rangle \hfill \\
  \left\langle {{T_{xx}}} \right\rangle  =& \frac{1}
{{4{G_N}}}\{ {h_{(2)xx}} + \frac{1}
{{2\lambda  + 1}}{h_{(2)tx}}\}  = \left\langle {{T_{tt}}} \right\rangle \hfill \\
\end{split}
\label{3.12}
\end{equation}
\begin{equation}
{h_{(2)}} = \frac{{{G_N}(2\lambda  + 1)}}
{{\lambda (\lambda  + 1)}}\left( {\begin{array}{*{20}{c}}
   {(2\lambda  + 1)\left\langle {{T_{tt}}} \right\rangle  - \left\langle {{T_{tx}}} \right\rangle } & { - \left\langle {{T_{tt}}} \right\rangle  + (2\lambda  + 1)\left\langle {{T_{tx}}} \right\rangle }  \\
   { - \left\langle {{T_{tt}}} \right\rangle  + (2\lambda  + 1)\left\langle {{T_{tx}}} \right\rangle } & {(2\lambda  + 1)\left\langle {{T_{tt}}} \right\rangle  - \left\langle {{T_{tx}}} \right\rangle }  \\
\end{array} } \right)
\label{3.13}
\end{equation}
\begin{equation}
\begin{split}
  \left\langle {{X_{tt}}} \right\rangle  =& \frac{{\lambda (1 + \lambda )}}
{{4{G_N}(2\lambda  + 1)}}\{ {h_{(2 + 2\lambda )}}_{tt} - {h_{(2 + 2\lambda )}}_{xt}\}  \hfill \\
  \left\langle {{X_{tx}}} \right\rangle  =& \frac{{\lambda (1 + \lambda )}}
{{4{G_N}(2\lambda  + 1)}}\{ {h_{(2 + 2\lambda )}}_{tx} - {h_{(2 + 2\lambda )}}_{xx}\}  =  - \left\langle {{X_{tt}}} \right\rangle  \hfill \\
  \left\langle {{X_{xt}}} \right\rangle  =& \frac{{\lambda (1 + \lambda )}}
{{4{G_N}(2\lambda  + 1)}}\{ {h_{(2 + 2\lambda )}}_{xt} - {h_{(2 + 2\lambda )}}_{tt}\}  =  - \left\langle {{X_{tt}}} \right\rangle  \hfill \\
  \left\langle {{X_{xx}}} \right\rangle  =& \frac{{\lambda (1 + \lambda )}}
{{4{G_N}(2\lambda  + 1)}}\{ {h_{(2 + 2\lambda )}}_{xx} - {h_{(2 + 2\lambda )}}_{tx}\}  = \left\langle {{X_{tt}}} \right\rangle  \hfill \\
\end{split}
\label{3.14}
\end{equation}
\begin{equation}
{h_{(2 + 2\lambda )}} = \frac{{{G_N}(2\lambda  + 1)}}
{{\lambda (\lambda  + 1)}} \cdot 2\left\langle {{X_{tt}}} \right\rangle \left( {\begin{array}{*{20}{c}}
   1 & { - 1}  \\
   { - 1} & 1  \\
\end{array} } \right)
\label{3.15}
\end{equation}

\bibliographystyle{utphys}

\providecommand{\href}[2]{#2}\begingroup\raggedright\endgroup

\end{document}